\documentclass[11pt,a4paper]{article}
\usepackage{cmath}
\usepackage{graphicx}
\usepackage{caption}
\usepackage{subcaption}
\usepackage{float}
\usepackage{amssymb,amsmath}
\title{Running vacuum cosmology with bulk viscous matter}
\author{Sarath N\footnote{sarath@cusat.ac.in} , Titus K. Mathew\footnote{titus@cusat.ac.in}   and Jerin Mohan N. D\footnote{jerinmohandk@cusat.ac.in} \\\small\textit{Department of Physics}\\\small\textit{Cochin University of Science and Technology}\\ \small\textit{Kochi-22, India}}
\date{}
\begin{document}
	\sloppy
	\maketitle
	\begin{abstract}
		We study the late accelerating expansion of the universe by incorporating bulk viscous matter with the running vacuum. The vacuum energy density varies as the squares of the Hubble parameter ($\rho_{\Lambda}\propto H^2$), and the coefficient of bulk viscosity of matter is proportional to the velocity of expansion ($\xi \propto H$). We have obtained an analytical solution to the Friedmann equations and estimated the model parameters by adopting the chi-square goodness-of-fit using the combined data set SN1a+CMB+BAO+OHD. Using the best-fit parameters, we have evaluated the universe's age as 14 Gyr, which is slightly higher than the age-predicted by the $\Lambda$CDM model. However, it is an improved result compared to the age-predicted by a class of bulk viscous matter-dominated models. Interestingly, we have obtained the coefficient of bulk viscosity of the matter component as $1.316\times 10^5$ kg $\textnormal m^{-1}$ $\textnormal s^{-1}$ which is one to two orders of magnitude less than the value predicted by most of the bulk viscous matter-dominated models and it falls in the range of highly viscous materials found on the earth. The Hubble parameter is a decreasing function of the scale factor, and it attains a constant value in the far future that corresponds to an end deSitter phase of evolution. The deceleration parameter shows a transition from matter-dominated decelerated phase to vacuum energy-dominated accelerating phase, and the transition redshift is obtained as $z_T = 0.73$. The statefinder analysis distinguishes our model from the $\Lambda$CDM model at present, and the $r-s$ trajectory reveals the quintessence behaviour of the vacuum energy. The phase space analysis shows that the universe is evolving towards a mechanically stable state in the far future. The entropy evolution satisfies the generalised second law of thermodynamics, and the entropy is maximised in the far future evolution. 
	\end{abstract}

	\section{Introduction}
	
The most successful model which explain the late time  cosmic acceleration\cite{riess1998observational,perlmutter1999constraining,spergel2003first,tegmark2004cosmological, carroll2001cosmological} is the $\Lambda$CDM model. This model assumes  
cosmological constant $\Lambda$ with 
equation of state $\omega_{\Lambda} = -1$ as the dark energy, 
which triggers the accelerated expansion\cite{carroll2001cosmological, carroll1992cosmological, padmanabhan2003cosmological, 1984ApJ...284..439P,peebles2003cosmological, frieman2008dark}.   
However, the cosmological constant predicted by
quantum field theory, as vacuum energy density, comes about 120 orders of magnitude above its observational value\cite{padmanabhan2003cosmological, weinberg1989cosmological, carroll1992cosmological}. 
This vast discrepancy, which is known as the cosmological constant problem\cite{weinberg1989cosmological}, does not have any explanation in the standard $\Lambda$CDM model. The standard model also failed to account for the coincidence of the current densities of dark matter and dark energy, though the evolutionary history of both these components is different\cite{dalal2001testing, schutzhold2002cosmological}.
A variety of alternate dark energy models have been proposed to circumvent these shortcomings. These models are realized by either modifying the curvature term or the energy-momentum term in the gravitational field equation. The former class of models are those devoid of any conventional dark energy, are generally known as modified gravity theories\cite{capozziello2002curvature, ferraro2007modified, dvali20004d, nojiri2005gauss, padmanabhan2013lanczos, amendola1999scaling, hovrava2009quantum}. The latter models include the dynamical dark energy and the dissipative cosmological models\cite{amendola2000coupled, tsujikawa2013quintessence, armendariz2001essentials, kamenshchik2001alternative, padmanabhan1987viscous, waga1986bulk, nair2016bulk}. Among these, the dynamical dark energy models deal with a dark energy density that varies as the universe expands. In contrast, dissipative models will generally invoke the dissipative effects like viscosity in the matter sector to generate the accelerated expansion. 

One of the dynamical dark energy models of recent interest is the running vacuum model 
\cite{tsiapi2019testing, peracaula2021running, peracaula2019signs}, which is proposed using the renormalization group theory. In these models, the dark energy candidate is a dynamically varying vacuum energy.
Many authors have studied this model extensively using the latest cosmological observational data and tested its significance against the standard $\Lambda$CDM model\cite{tsiapi2019testing, peracaula2021running, peracaula2019signs, sola2017first, sola2018possible, zhao2017dynamical}. For instance, in previous work, we have confronted this model with the observational data using Bayesian statistics and found a definite, but not strong, evidence for the running vacuum model against the standard $\Lambda$CDM\cite{10.1093/mnras/stab3773}.

This paper analyses the running vacuum model of the late accelerating universe with the bulk viscous matter, and hence the model can be abbreviated as RVBVM. In this model, the dark energy is the running vacuum\cite{basilakos2013inflation}, while the matter component is assumed to have a bulk viscosity. We use Eckart formalism to incorporate bulk viscous pressure with the matter sector\cite{PhysRev.58.919}. Following this, the effective pressure generated due to bulk viscosity amounts to $-3\xi H,$ where $H$ is the Hubble parameter. In literature, we have models describing the late acceleration using bulk viscous matter alone as a cosmic component, just like running vacuum models, which explain the late acceleration using varying vacuum energy. Both these types of models are aimed to alleviate the major drawbacks of the standard $\Lambda$CDM. Among these, we have already mentioned that the significance of the running vacuum model is not so strong against the standard model. The same is the case with the viscous model, which will be described in the following paragraph. It is this motivated us to consider the present model, which combined the effect of both running vacuum and bulk viscous matter.

There are various models in the literature which are explaining the late acceleration with bulk viscous matter alone as a cosmic component\cite{cataldo2005viscous, hu2006bulk, israel1979transient, avelino2013bulk, fabris2006bulk, kremer2003viscous, mohan2017bulk, normann2016general, normann2017characteristic, sasidharan2015bulk}. These models reasonably explain the recent acceleration. In ref. \cite{sasidharan2016phase}, the authors performed a dynamical system analysis and thermal evolution of various bulk viscous models based on Eckart formalism and show that these models are dynamically stable and thermodynamically consistent. However, the analysis based on Bayesian inference presented in \cite{sasidharan2018bayesian} shows that the evidence of these types of bulk viscous models against the standard $\Lambda$CDM model is weak. An extensive analysis of full causal bulk viscous models and their truncated version are presented in ref.\cite{mohan2021feasibility, mohan2017bulk}. These models also provide a good background evolution, and the models are also dynamically stable and thermodynamically consistent\cite{JerinMohan:2020qqp, mohan2021feasibility}. A very recent analysis based on Bayesian statistics shows weak evidence of the causal dissipative model against the standard $\Lambda$CDM model. Thus, it is worth incorporating the matter having bulk viscous pressure with the running vacuum to see to what extent the model could give a better description of the late accelerating epoch. 

The paper is systematized as follows. In section 2, we obtain an analytical solution to the Friedmann equation. In section 3, We perform chi-square goodness-of-fit to estimate the model parameters. Section 4 presents the evolution of the cosmological parameters. In Section 5, Statefinder analysis is performed to differentiate our model from the $\Lambda$CDM model. In section 6, we present a dynamical system analysis to explain the mechanical stability of the model. In section 7, we carry out the thermodynamical stability analysis and check the validity of the generalized second law of thermodynamics. Finally, we conclude in section 8.

	\section{Running vacuum with Bulk Viscous Matter}
	Modern cosmological observations indicate that our universe is spatially flat and obeys the cosmological principle. Such a universe is well described by an exact solution of Einstein's field equation known as Friedmann–Lemaitre–Robertson–Walker (FLRW) metric. In flat space, the metric is
	\begin{equation}
	\label{eqn:metric}
	ds^2 = -dt^2 + a(t)^2(dr^2 + r^2d\theta^2 + r^2\sin^2(\theta)d\phi^2),
	\end{equation}
	where $a(t)$ is the scale factor that parametrize the relative expansion of the universe, ($r, \theta, \phi$) are the co-moving coordinates and $t$ is the cosmic time. The evolution of the universe is characterized by Friedmann equations that are derived from gravitational field equations by assuming FLRW metric is
	\begin{equation}
	\label{eqn:F01}
	3H^2 = 8\pi G(\rho_m + \rho_{\Lambda}),
	\end{equation}
	\begin{equation}
	\label{eqn:F02}
	2\dot{H} + 3H^2 = -8\pi G(\text{p}_{\text{eff}} + \text{p}_{\Lambda}),
	\end{equation}
	where $H=\frac{\dot{a}}{a}$ is the Hubble parameter, overdot represents the derivative with respect to cosmic time. The vacuum energy density ($\rho_{\Lambda}$) and matter density ($\rho_m$) satisfies the common conservation equation
	\begin{equation}
	\label{eqn:c1}
	\dot{\rho_m}+\dot{\rho_{\Lambda}}+3H(\rho_m+\text{p}_{\text{eff}})+3H(\rho_{\Lambda}+p_{\Lambda})=0.
	\end{equation}
	Here, we focuss on the late accelerating epoch; the radiation density and hence the radiation pressure ($p_{\gamma}$) are negligibly small and can be ignored. We consider nonrelativistic matter having the density $\rho_m$ and effective pressure of the form\cite{ cataldo2005viscous, Sasidharan:2014wqa, weinberg1972gravitation},
	\begin{equation}
	\label{eqn:p}
	\text{p}_{\text{eff}} = P-3\xi H, 
	\end{equation}
	where P is the normal pressure which is zero for nonrelativistic matter and $-3\xi H$ is the bulk viscous pressure. The coefficient of bulk viscosity $\xi$, in general, can depend on the velocity and acceleration of expansion. However, we consider the $\xi$ depends only on the velocity of expansion to get analytical solution feasible, and hence it is expressed as\cite{normann2016general}
	\begin{equation}
	\label{eqn:sy}
	\xi =\xi_1 H.
	\end{equation}
	We consider vacuum energy as a perfect fluid with equation of state $\omega_{\Lambda}= -1$. Hence the vacuum pressure is just the vacuum energy density with a negative sign ($p_{\Lambda} = -\rho_{\Lambda}$). Instead of considering vacuum energy as a cosmological constant, we consider it as a dynamical entity and that is perfectly allowed by renormalization group equation of quantum field theory in curved spacetime geometry,
	\begin{equation}
	\label{eqn:r1}
	\frac{d\rho_{\Lambda}(\mu)}{d\ln \mu^2} = \frac{1}{(4\pi^2)}\sum_{i}\left[a_iM_i^2\mu^2 + b_i\mu^4 + c_i\frac{\mu^6}{M_i^2} + ...\right],
	\end{equation}
	where the $\rho_{\Lambda}$ is a function of renormalization running scale $\mu$, which is naturally associated with the Hubble parameter and its time derivatives. The coefficients $a_i$, $b_i$, $c_i$ are dimensionless constants and $M_i$ are the massess of the fermionic and bosonic particles contributing the loop. The local covariance of effective action allows only the even powers of the Hubble parameter. The dynamics of the present universe is encoded in the $H^2$ term while the higher order terms are needed to explain the very rapid expansion in the early inflationary epoch. Note that the $M_i^4$ terms are absent because it triggers very fast running of vacuum energy. Moreover, these possibilities are excluded since the field satisfying $\mu > M_i$ are the active degrees of freedom in the RG formulation\cite{sola2018tensions}. The vacuum energy density that depends on the Hubble parameter is expressed as
	\begin{equation}
	\label{eqn:rho1}
	\rho_{\Lambda}(H)=\frac{3}{8\pi G}\left(c_0 + \nu H^2\right),
	\end{equation}
	where $c_0\sim \Lambda/3$, $\Lambda$ is the cosmological constant. The dimensionless coefficient $\nu$; depends on the square of masses of the matter fields plays the role similar to that of the $\beta$-function coefficient of effective action of QFT in curved geometry,
	\begin{equation}
	\label{eqn:nu}
	\nu = \frac{1}{6\pi}\sum_{i=f,b}^{}B_i\frac{M_i^2}{M_P^2}.
	\end{equation}
	Substituting the expression of bulk viscous pressure $p_m$ from eq. (\ref{eqn:p}) and the vacuum pressure $p_{\Lambda} = -\rho_{\Lambda}$ in eq. (\ref{eqn:F02}), we obtain
	\begin{equation}
	\label{eqn:f03}
	\dot{H}=\frac{3}{2}c_0-\frac{3}{2}(1-\nu-\tilde\xi_1)H^2,
	\end{equation}
	where $\tilde\xi_1=8\pi G\xi_1$ is the coefficient of bulk viscosity expressed as dimensionless. We consider a unit in which $8\pi G=1$ for the further discussion. It is appropriate to change integration variable from time to scale factor, we obtain 
	\begin{equation}
	\label{eqn:f04}
	\frac{HdH}{(1-\nu-\tilde\xi_1)H^2-c_0}=-\frac{3da}{2a}.
	\end{equation}
	On integration, the Hubble parameter evolution in a universe dominated by vacuum energy and the bulk viscous matter is obtained as
	\begin{equation}
	\label{eqn:H}
	H=\left[\left(H_0^2-\frac{c_0}{1-\nu-\tilde\xi_1}\right)a^{-3(1-\nu-\tilde\xi_1)}+\frac{c_0}{1-\nu-\tilde\xi_1}\right]^{\frac{1}{2}}.
	\end{equation}
	It is convenient to express the Hubble parameter evolution presented in eq. (\ref{eqn:H}) in terms of observables such as matter density parameter $\Omega_{m_0}$ and the redshift $z$, we have 
	\begin{equation}
	\label{eqn:H1}
	H=H_0\left[\left(\frac{\Omega_{m_0}-\tilde\xi_1}{1-\nu-\tilde\xi_1}\right)(1+z)^{3(1-\nu-\tilde\xi_1)}+\frac{1-\nu-\Omega_{m_0}}{1-\nu-\tilde\xi_1}\right]^{\frac{1}{2}},
	\end{equation}
	where we have substituted for the scale factor $a =1/(1+z)$ and $c_0 = (1-\nu-\Omega_{m_0})$. Here $\Omega_{m_0}$ is the matter density parameter at present, which is related to the present matter density $\rho_{m_0}$ as $\Omega_{m_0}=\rho_{m_0}/\rho_c$, where $\rho_c=3H_0^2$ is the critical density of the universe.
	From eq. (\ref{eqn:H1}), it is clear that the model reduces to running vacuum model when $\tilde\xi_1=0$, and it reduces to $\Lambda$CDM model when $\tilde\xi_1=0$ and $\nu=0$. When $z=0$ (equivalently $a=1$), the eq. (\ref{eqn:H1}) reduces to $H=H_0$ where $H_0$ is the Hubble parameter at present. In the asymptotic limit $z\rightarrow -1$ (equivalently $a\rightarrow\infty$), $H\rightarrow H_0[(1-\nu-\Omega_{m_0})/(1-\nu-\tilde\xi_1)]^{1/2}$ is constant, that corresponds to an end deSitter phase  of evolution.  
	
	\section{Parameter Estimation}
	The model possess five unknown parameters, $\nu$, $\tilde{\xi_1}$, $H_0$, $\Omega_{m_0}$ and $\Omega_{\Lambda_0}$. However, the matter density parameter $\Omega_{m_0}$ and the vacuum density parameter $\Omega_{\Lambda_0}$ are not independent, constrained by the relation $\Omega_{m_0} + \Omega_{\Lambda_0} = 1$, in a Flat FLRW universe\cite{peebles1993principles,liddle2015introduction, weinberg1973gravitation}. Hence the number of independent model parameters reduces to four. We estimate the model parameters using the $\chi^2$ minimization technique in the light of observational data. Here, we choose the combined cosmological observations, SNIa+CMB+BAO+OHD, to obtain the goodness-of-fit. The largest spectroscopically confirmed Type Ia supernovae data set is the pantheon sample that contains 1048  apparent magnitude versus redshift data in the redshift range $0.01< z < 2.3$\cite{scolnic2018complete}. The sample set is a collection of 279 type Ia supernovae located by the Pan-STARRS1 medium-deep survey, the distance measurement from the
	Sloan Digital Sky Survey (SDSS), Supernova Legacy Survey (SNLS) and from diverse low
	redshift and Hubble Space Telescope samples.  To compute $\chi^2$, we compare the observed apparent magnitude with the theoretical one computed from the model. The luminosity distance $d_L$ of the $i^{th}$ supernovae with redshift $z_i$ in a flat universe is defined as
	\begin{equation}
	d_L(\nu, \tilde{\xi_1}, H_0, \Omega_{m_0},z_i) = c(1+z_i)\int_{0}^{z_i}\frac{dz}{H(\nu, \tilde{\xi_1}, H_0, \Omega_{m_0},z_i)},
	\end{equation}
	where $H (\nu, \tilde{\xi_1}, H_0, \Omega_{m_0},z_i)$ is the Hubble parameter measured in $\textrm{km} \textrm{s}^{-1}\textrm{Mpc}^{-1}$ and c is the speed of light in vacuum expressed in $\textrm{km} \textrm{s}^{-1}$ \cite{sasidharan2015bulk, george2019interacting}. The theoretical apparent magnitude  is estimated using the expression
	\begin{equation}
	\label{eqn:app}
	m(\nu, \tilde{\xi_1}, H_0, \Omega_{m_0},z_i) = 5\log_{10}\left[\frac{d_L(\nu, \tilde{\xi_1}, H_0, \Omega_{m_0},z_i)}{Mpc}\right] + 25 + M,
	\end{equation}
	where $M$ is a nuisance parameter, absolute magnitude of the type Ia supernova. We compare the apparent magnitude of type 1a supernovae with the corresponding theoretical one to obtain the $\chi^2_{SNIa}$. 
	We use shift parameter ($\mathcal{R}$) from Planck 2018 Cosmic Microwave Background (CMB) anisotropy to obtain $\chi_{CMB}^2$ . The shift parameter is usually derived from the position of the first acoustic peak in the power spectrum of CMB temperature anisotropies assuming a spatially flat universe with dark matter and cosmological constant\cite{elgaroy2007using}. The observed shift parameter is $\mathcal{R}_{obs}=1.7502\pm0.0046$ at a redshift of $z_1 = 1089.92$\cite{chen2019distance, calderon2021negative, capistrano2020subhorizon, rivera2019exploring}. The shift parameter can be estimated theoretically as,
	\begin{equation}
	\label{eqn:shift}
	\mathcal{R}_t = \sqrt{\Omega_{m_0}}\int_{0}^{z_1}\frac{dz'}{h(z')},
	\end{equation}
	where $h=H/H_0$ is the reduced Hubble parameter.
	The Baryonic Acoustic Oscillation (BAO) is the periodic fluctuations in the density field, stamped in the primordial plasma before decoupling. We use BAO data obtained from SDSS having the acoustic parameter $\mathcal{A} = 0.484\pm0.016$ at a redshift $z_1 = 0.35$ to obtain the $\chi_{BAO}^2$\cite{blake2011wigglez}. The theoretical acoustic parameter corresponds to a redshift $z_k$ is expressed as
	\begin{equation}
	\label{eqn:acou}
	\mathcal{A}_t = \frac{\sqrt{\Omega_{m_0}}}{h(z_1)^\frac{1}{3}}\left(\frac{1}{z_1}\int_{0}^{z_1}\frac{dz}{h(z)}\right)^\frac{2}{3},
	\end{equation}
	where $z_k$ is the redshift where the signature of peak acoustic oscillation is measured. 
	We use the Observational Hubble Data (OHD) having 36 data points in the redshift range  $0.0708\leq z \leq 2.36$ to obtain $\chi_{OHD}^2$\cite{amirhashchi2020constraining, moresco2015raising, rivera2019exploring}. The 26 data points were obtained from the differential age method, and the remaining 10 data were procured from the radial BAO method. 
	Then, the $\chi^2$ corresponding to each data set can be calculated using the expression.
	\begin{equation}
	\label{eqn:chisquare}
	\chi^2(\nu, \tilde{\xi_1}, H_0, \Omega_{m_0}) = \sum_{k}\frac{\left[\mathcal{E}(\nu, \tilde{\xi_1}, H_0, \Omega_{m_0},z_k)-\mathcal{O}_{k}\right]^2}{\sigma_k^2},
	\end{equation}
	where, $\mathcal{O}_k$ is the physicsl quantity obtained from the data at redshift $z_k$, $\mathcal{E}$ is the corresponding physical quantity obtained from the model and $\sigma_k$ is the corresponding standard deviation in the measurement. 
	The parameters that best fit the data combination SNIa+CMB+BAO+OHD is obtained by minimizing the $\chi^2$ of the form
	\begin{equation}
	\label{eqn:chi}
	\chi^2_{total} = \chi^2_{SNIa}+\chi^2_{CMB}+\chi^2_{BAO}+\chi^2_{OHD}
	\end{equation}
	We have obtained the absolute magnitude of the type Ia supernovae, $M = -19.3963$, by minimizing the combined $\chi^2$ given in eq. (\ref{eqn:chi}). 
	\begin{figure}[H]
		\begin{subfigure}{.5\textwidth}
			\includegraphics[width=0.86\linewidth]{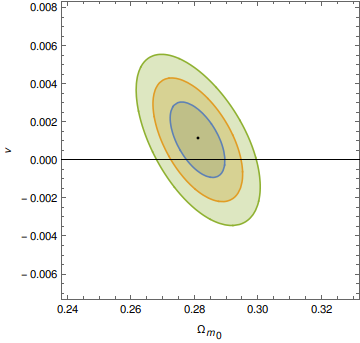}
			\caption{}
			\label{fig:omnu}
		\end{subfigure}
		\begin{subfigure}{.5\textwidth}
			\includegraphics[width=0.87\linewidth]{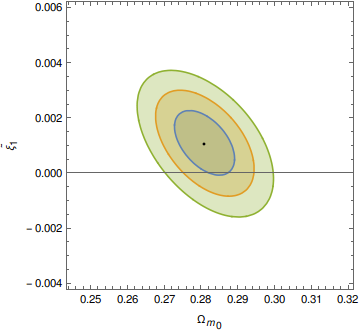}
			\caption{}
			\label{fig:omH0}
		\end{subfigure}
		\caption{Confidence contour of the parameters (a) $(\Omega_{m_0}, \nu)$ and (b) $(\Omega_{m_0}, \tilde{\xi_1})$ using the SNIa+CMB+BAO+OHD datasets. The dot indicate the best fit values of the respective parameters.  The confidence contours shown are corresponds to the 68.3\%, 95.4\% and 99.73\% of probabilities.}
		\label{fig:contour}
	\end{figure}
	\noindent We have constructed the $1\sigma(68.3\%)$, $2\sigma(95.4\%)$, and $3\sigma(99.73\%)$ confidence contour of $(\Omega_{m_0},\nu)$ and $(\Omega_{m_0},\tilde\xi_1)$ as shown in figure (\ref{fig:contour}). 
	The parameter values that best describe the evolution of the universe within $1\sigma$ confidence are, $H_0=68.7970\pm0.2400$, $\Omega_{m_0}=0.2809\pm0.0089$, $\tilde{\xi_1}=0.0011^{+0.0012}_{-0.0011}$ and $\nu=0.0011^{+0.0020}_{-0.0011}$. The estimated value of the parameter $\nu$ is almost similar to the value obtained for running vacuum models\cite{tsiapi2019testing, sola2015hints, gomez2015dynamical}. The most interesting part of the calculation is the estimated value of $\tilde{\xi}_1$ which is atleast two orders of magnitude less as compared to the value obtained for bulk viscous models presented in ref. \cite{sasidharan2015bulk, Mohan:2017poq, sasidharan2018bayesian}, where sasidharan et al. has obtained $\tilde{\xi}_1 = 1.683.$
	\section{Evolution of Cosmological Parameters}
\begin{figure}[H]
		\centering
		\includegraphics[width=0.5\linewidth]{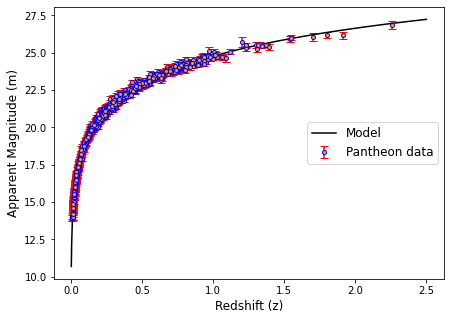}
		\caption{Comparison between the observed apparent magnitude of type Ia supernovae within the redshift range $0.01< z< 2.3$ (cyan scattered points) and the apparent magnitude predicted by the decaying vacuum with bulk viscous matter (solid black line). The error bars corresponds to the observational data.}
		\label{fig:app}
	\end{figure}

\begin{figure}[H]
		\centering
		\includegraphics[width=0.5\linewidth]{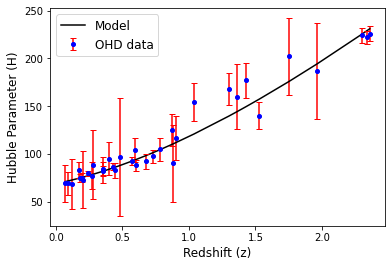}
		\caption{Hubble parameter (H) evolution of decaying vacuum model with bulk viscous matter (solid black line) is plotted againist redshift (z). The scattered points are the Hubble parameter values in the redshift range $0.0708\leq z \leq 2.36$ from observational Hubble data and the vertical lines represents the standard deviation in the measured values.}
		\label{fig:H}
	\end{figure}

	The apparent magnitude of type 1a supernovae determined using the model is presented in Eq. (\ref{eqn:app}). The pantheon data set contain 1048 apparent magnitude of type 1a supernovae measured at different redshifts. In fig. (\ref{fig:app}), we compare the theoretical apparent magnitude obtained using the best-fit parameters and the observed apparent magnitude of type 1a supernovae. The model is in good agreement with the type 1a supernovae data. 
	\noindent The Hubble parameter evolution given in eq. (\ref{eqn:H1}) is plotted against the redshift as shown in fig. (\ref{fig:H}), which in this model fits quite well with the OHD data at low redshift. At high redshift, the model seems to be deviating from the observational data.
	
	The age of the universe can be estimated as follows. The Hubble parameter is defined as $H(a)=\dot{a}/a$, rearranging we obtain $da/dt=aH(a)$ or $dt/da=(aH(a))^{-1}$. Then, the age of the universe is expressed as
	\begin{equation}
	\label{eqn:age}
	t_0-t_B = \int_{0}^{1}\frac{1}{aH(a)}da,
	\end{equation}
	where $t_0$ is the present time, and $t_B$ is the time corresponding to the big bang. Since the evolution of the universe is assumed to be started from the big bang, $t_B=0$, then $t_0$ represents the age of the universe. Substituting the Hubble parameter expression from eq. (\ref{eqn:H1}), we obtain the age of the universe 14 Gly, which is slightly higher than the age, 13.74 Gyr obtained from the CMB anisotropy data\cite{tegmark2006cosmological}, $12.9\pm2.9$Gyr, measured from the oldest globular clusters\cite{carretta2000distances} and $\approx$13.8 Gyr, the age-predicted by $\Lambda$CDM model\cite{weinberg1972gravitation}. Nevertheless, this is quite an improved result compared to the relatively low age, 10.90 Gyr predicted by the non-causal bulk viscous model\cite{sasidharan2015bulk}, 9.72 Gyr estimated from the causal viscous model\cite{Mohan:2017poq} and 10–12
	Gyr obtained using viscous Zel’dovich fluid models\cite{Nair:2015bhz}. 
	
\begin{figure}[H]
		\centering
		\includegraphics[width=0.5\linewidth]{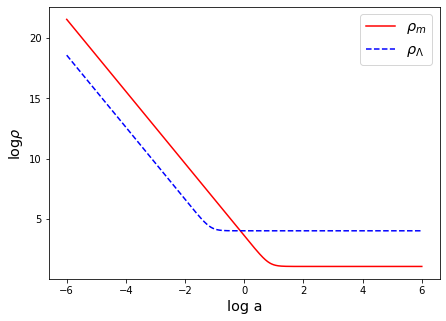}
		\caption{Progress of both matter density ($\rho_m$) and vacuum energy density ($\rho_{\Lambda}$) is plotted against scale factor (a) in logarithemic scale.}
		\label{fig:D}
	\end{figure}

	\noindent The evolution of vacuum energy density is given in eq. (\ref{eqn:rho1}). Substituting the eq. (\ref{eqn:H1}) in (\ref{eqn:rho1}), we obtain the expression for vacuum energy density as
	\begin{equation}
	\label{eqn:rho2}
	\rho_{\Lambda} = \rho_{\Lambda_0}+\frac{\nu}{1-\nu-\tilde{\xi_1}}\left[(\rho_{m_0}+\text{p}_{\text{eff}}^0)\left(a^{-3(1-\nu-\tilde{\xi_1})}-1\right)\right],
	\end{equation}
	where $\rho_{\Lambda_0}$, $\rho_{m_0}$ and $\text{p}_{\text{eff}}^0$ are the present value of the vacuum energy denisty, matter density and bulk viscous pressure respectevely. When $a=1$, $\rho_{\Lambda} = \rho_{\Lambda_0}.$ In the asymptotic limit, $a\rightarrow \infty$, the vacuum energy density become a constant, $\rho_{\Lambda} = \rho_{\Lambda_0}-\frac{\nu}{1-\nu-\tilde{\xi_1}}\left(\rho_{m_0}+\text{p}_{\text{eff}}^0\right)$. In the absence of bulk viscous matter $\tilde{\xi_1} =0$ and hence $\text{p}_{\text{eff}}^0=0$, the form of vacuum energy density reduces to that of the running vacuum model. Furthermore when $\nu=0$, the vacuum energy density is just the cosmological constant density $\rho_{\Lambda_0}$. 
	Substituting the expression of vacuum energy density presented in eq. (\ref{eqn:rho2}) and the Hubble parameter expression from eq. (\ref{eqn:H1}), we obtain the evolution of matter density as 
	\begin{equation}
	\label{eqn:rho3}
	\rho_m = \frac{(1-\nu)(\rho_{m_0}+\text{p}_{\text{eff}}^0)}{1-\nu-\tilde{\xi_1}}a^{-3(1-\nu-\tilde{\xi_1})}+\frac{\tilde{\xi_1}\rho_{\Lambda_0}}{1-\nu-\tilde{\xi_1}}+
	\frac{\nu \text{p}_{\text{eff}}^0}{1-\nu-\tilde{\xi_1}}.
	\end{equation}
	The present matter density $\rho_m=\rho_{m_0}$ is obtained by substituting $a=1$ in eq. (\ref{eqn:rho3}). In the asymptotic limit, $a\rightarrow \infty$, the matter energy density become a constant, $\rho_{m} = \frac{\tilde{\xi_1}\rho_{\Lambda_0}}{1-\nu-\tilde{\xi_1}}+
	\frac{\nu \text{p}_{\text{eff}}^0}{1-\nu-\tilde{\xi_1}}$.
	When $\tilde{\xi_1}=0$, the pressure due to bulk viscous matter $\text{p}_{\text{eff}}^0=0$, then the matter density reduces to that of the running vacuum model $\rho_m = \rho_{m_0}a^{-3(1-\nu)}$ and it further reduces to $\rho_m=\rho_{m_0}a^{-3}$ when $\nu=0$, which corresponds to matter density evoution in $\Lambda$CDM model. Evolution of both matter density and vacuum energy density as a function of scale factor in logarithemic scale is plotted in fig. (\ref{fig:D}).
	\noindent From the figure, it is conclusive that the gravitating matter is the dominant component up to the recent past. It represents a decelerated expansion of the universe, the vacuum energy dominated over the matter density in the recent past and triggered the accelerating expansion of the universe. The time evolution of the vacuum gives an effortless explanation to the cosmic coincidence problem. 
	
	The bulk viscous pressure is given by eq. (\ref{eqn:p}). Substituting for the coeficient of bulk viscosity from eq. (\ref{eqn:sy}), we obtain 
	\begin{equation}
	\label{eqn:p1}
	\text{p}_{\text{eff}} = -3\tilde{\xi_1} H^2.
	\end{equation}
	The present coefficient of bulk viscosity is obtained for the best fit parameter $\tilde{\xi_1}=0.0011$ and $H_0=68.7970$, the estimated value is $1.316\times 10^5$ kg $\textnormal m^{-1}$ $\textnormal s^{-1}$. The coefficient of bulk viscosity estimated by many bulk viscous model is of the order of $10^6 - 10^7$ \cite{Sasidharan:2014wqa, Mohan:2017poq, sasidharan2018bayesian} which is one to two orders of magnitude higher as compared to the highly viscous materials found on the earth. The incorporation of bulk viscous matter with the decaying vacuum is successful in contraining the bulk viscosity two orders of magnitude below the value predicted by matter dominated bulk viscous models. Substituting the expression of the Hubble parameter from the eq. (\ref{eqn:H1}), we obtain the evoltion of the bulk viscous pressure as
	\begin{equation}
	\label{eqn:p2}
	\text{p}_{\text{eff}} = \text{p}_{\text{eff}}^0\left[\frac{\Omega_{m_0}-\tilde{\xi_1}}{1-\nu-\tilde{\xi_1}}a^{-3(1-\nu-\tilde{\xi_1})}+\frac{1-\Omega_{m_0}-\nu}{1-\nu-\tilde{\xi_1}}\right].
	\end{equation}
	At present ($a=1$), the bulk viscous pressure $\text{p}_{\text{eff}} = \text{p}_{\text{eff}}^0$. Since the present value of the viscous pressure depends on $\tilde{\xi_1},$ at zero viscosity this represents a pressureless matter.
	
\begin{figure}[H]
		\centering
		\includegraphics[width=0.5\linewidth]{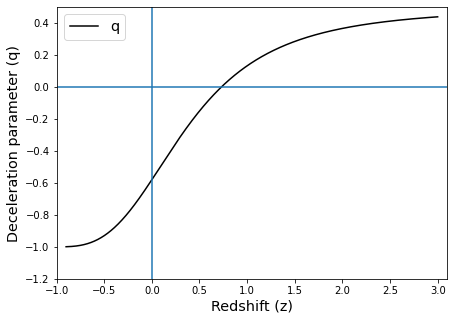}
		\caption{Evolution of deceleration parameter (q) is plotted against redshift (z).}
		\label{fig:d}
	\end{figure}

	The decceration parameter (q) is a dimensionless parameter that characterizes the accelerating or deccelerating expansion of the FLRW universe. It is defined as
	\begin{equation}
	\label{eqn:q}
	q = -1-\frac{\dot{H}}{H^2}.
	\end{equation}
	It is convenient to express the eq. (\ref{eqn:H1}) in terms of dimensionless Hubble parameter,
	\begin{equation}
	\label{eqn:q0}
	q = -1-\frac{1}{2h^2}\frac{dh^2}{dx},
	\end{equation}
	where $x=\ln(a)$. Substituting for $h^2=H^2/H_0^2$ from eq. (\ref{eqn:H1}), we obtain the deceleration parameter that depends on scale factor as
	\begin{equation}
	\label{eqn:q1}
	q=-1+\frac{3(\Omega_{m_0}-\tilde{\xi_1})a^{-3(1-\nu-\tilde{\xi_1})}}{2\left[\left(\frac{\Omega_{m_0}-\tilde{\xi_1}}{1-\nu-\tilde{\xi_1}}\right)a^{-3(1-\nu-\tilde{\xi_1})}+\frac{1-\nu-\Omega_{m_0}}{1-\nu-\tilde{\xi_1}}\right]}.
	\end{equation}
	The evolution of deceleration parameter with respect to redshift (z) is plotted in fig. (\ref{fig:d}). 
	\noindent The universe made a smooth transition from deccelerated phase to accelerating phase at a transition redshift $z_T=0.7306$ that is comparable to the transition redshift, $Z_T=0.50-0.73$ obatined for the $\Lambda$CDM model\cite{alam2004case,colistete2007bulk}. The current decceleration parameter value is obtained by substituting $a=1$ (or equivalently $z=0$) in eq. (\ref{eqn:q1}),
	\begin{equation}
	\label{eqn:q2}
	q_0=-1+\frac{3}{2}(\Omega_{m_0}-\tilde{\xi_1})
	\end{equation}
	Interestingly, the present deceleration parameter is independent of the parameter $\nu$. The expression of $q_0$ reduces to that of the $\Lambda$CDM model when $\tilde{\xi_1}=0$. Using the best-estimated value of the model parameter, the estimated value of the deceleration parameter is $q_0=-0.5802$, which is close to the value predicted by the $\Lambda$CDM model\cite{liddle2015introduction}. 
	\section{Statefinder Analysis}
\begin{figure}[H]
		\centering
		\includegraphics[width=0.5\linewidth]{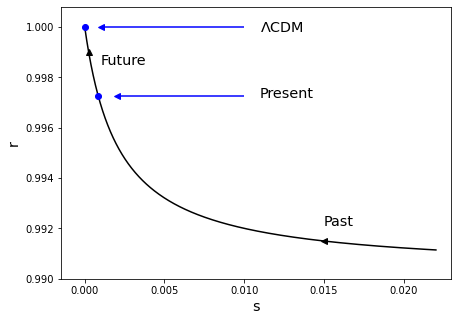}
		\caption{r-s plot for best estimated model parameters.}
		\label{fig:rs}
	\end{figure}

	Statefinder is a geometric diagnostic tool introduced by Sahni et al.\cite{sahni2003statefinder} to distinguish between two dark energy models. The statefinder parameters \{$r$, $s$\} is used to probe the charecteristics of the various dark energy models. The statefinder pair \{$r$, $s$\} = \{1, 0\}, is a fixed point for the standard $\Lambda$CDM model. The jerk parameter (r) is followed by the deceleration parameter in the taylor series expansion of the scale factor and the snap parameter $s$ can be constructed from the r parameter and the deceleration parameter as
	\begin{equation}
	\begin{aligned}
	\label{eqn:s1}
	&r=\frac{\dddot{a}}{aH^3}, \\ 
	&s=\frac{r-1}{3(q-\frac{1}{2})},
	\end{aligned}
	\end{equation}
	where $\dddot{a}$ is the third derivative of scale factor with respect to cosmic time, $H$ is the Hubble parameter and $q$ is the deceleration parameter. Use of geometrical variables such as $H_0$, $q_0$, r and s has some advantages when characterizing the propertis of dark energy. The geometrical variables are more universal as they are constructed from the spacetime metric directly whereas the physical variables such as density is model dependent and are always measured with uncertainity\cite{sahni2003statefinder}. It is possible to express the $r$ and $s$ parameter in terms of reduced Hubble parameter $h=H/H_0$ as
	\begin{align}
	\label{eqn:s2}
	r= \frac{1}{2h^2}\frac{d^2h^2}{dx^2}+\frac{3}{2h^2}\frac{dh^2}{dx}+1,
	\end{align}
	\begin{align}
	\label{eqn:s3}
	s= -\frac{\frac{1}{2h^2}\frac{d^2h^2}{dx^2}+\frac{3}{2h^2}\frac{dh^2}{dx}}{\frac{3}{2h^2}\frac{dh^2}{dx}+\frac{9}{2}},
	\end{align}
	where $x=\ln a$\cite{Mathew:2012md}. Substituting the expression of $h^2$ given in eq.(\ref{eqn:H1}) in eq. (\ref{eqn:s2}) and (\ref{eqn:s3}), we obtain the expression for jerk parameter
	\begin{align}
	\label{eqn:s4}
	&r= 1-\frac{9(1-\nu-\tilde{\xi_1})(\Omega_{m_0}-\tilde{\xi_1})(\nu+\tilde{\xi_1})a^{-3(1-nu-\tilde{\xi_1})}}{2(\Omega_{m_0}-\tilde{\xi_1})a^{-3(1-\nu-\tilde{\xi_1})}+1-\nu-\Omega_{m_0}},
	\end{align}
	and the snap parameter
	\begin{align}
	\label{eqn:s5}
	&s= \frac{(1-\nu-\tilde{\xi_1})(\nu+\tilde{\xi_1})(\Omega_{m_0}-\tilde{\xi_1})a^{-3(1-\nu-\tilde{\xi_1})}}{(\nu+\tilde{\xi_1})a^{-3(1-\nu-\tilde{\xi_1})}+1-\nu-\Omega_{m_0}}.
	\end{align}
	The $r$ and $s$ parameter is $1$ and $0$ respectevely for the $\Lambda$CDM model throughout the evolution of the universe. In this model, we obtain the $r$ and $s$ parameters that depends on the scale factor. From eq. (\ref{eqn:s4}) and (\ref{eqn:s5}), it is evident that the parameters approches the $\Lambda$CDM model in the limit $a\rightarrow\infty$. The present $r$ and $s$ parameters are obtained by substituting $a=1$ in eq. (\ref{eqn:s4}) and (\ref{eqn:s5}), we obtain ($r_0$, $s_0$) = ($0.9972$, $0.0008$), shows that our model is obviously different from $\Lambda$CDM model. The ($r$,$s$) plot is shown in fig.(\ref{fig:rs}).
	\noindent The $r-s$ plot shows that $r<1$ and $s>0$ untill it reaches $\Lambda$CDM model, the model resembles quintessence dark energy models.
	\section{Phase Space Analysis}
	Dynamical system analysis is a useful technique to understand the asymtotic behaviour of a cosmological model when an exact solution to cosmological field equations are hardly possible. Here we perform the phase space analysis of the model to extract its asymtotic bahavior. Primarly we choose appropriate dynamical variables for the analysis. Here we consider dimensionless dynamical variables $u$ and $v$ which are defined as
	\begin{align}
	\label{eqn:d1}
	u=\frac{\rho_{\Lambda}}{3H^2},\hspace{0.5cm} v=\frac{\rho_{m}}{3H^2},
	\end{align}
	to set up autonomous coupled differential equations. On differentiating $u$ and $v$ presented in eq. (\ref{eqn:d1}) with respect to $x$, where x=$\ln a$,  we obtain
	\begin{align}
	\label{eqn:d2}
	\frac{du}{dx}=\frac{(\nu-u)}{H^2}\frac{dH^2}{dx},
	\end{align}
	\begin{align}
	\label{eqn:d3}
	\frac{dv}{dx}=\frac{(1-\nu-v)}{H^2}\frac{dH^2}{dx}.
	\end{align}
	Using the Freidmann equation presented in eq. (\ref{eqn:F02}), substituting for $dH^2/dx=-\rho_m + 3\tilde{\xi_1}H^2$ in eq. (\ref{eqn:d2}) and (\ref{eqn:d3}), we obtain
	\begin{align}
	\label{eqn:d4}
	\frac{du}{dx}=3(1-\nu-v)(1-u)-3\nu(1-u-\tilde{\xi_1}) = f(u, v),
	\end{align}
	\begin{align}
	\label{eqn:d5}
	\frac{dv}{dx}=3(v-\tilde{\xi_1})(\nu-u) = g(u, v).
	\end{align}
	The critical points ($u_c$ and $v_c$) are obtained by equating $\frac{du}{dx} = 0$ and $\frac{dv}{dx} = 0$. The stability of the universe in the vicinity of the critical points are obtained by considering a linear perturbation around the critical point, $u\rightarrow u' = u_c+\delta u$ and $v\rightarrow v' = v_c+\delta v$, we obtain the matrix equation
	\begin{equation}
	\label{eqn:M1}
	\begin{bmatrix}
	\delta u'\\
	\delta v'
	\end{bmatrix}
	=
	\begin{bmatrix}
	\left(\frac{\partial f}{\partial u}\right)_c & \left(\frac{\partial f}{\partial v}\right)_c \vspace{0.2cm}\\
	\left(\frac{\partial g}{\partial u}\right)_c & \left(\frac{\partial g}{\partial v}\right)_c
	\end{bmatrix}
	\begin{bmatrix}
	\delta u\\
	\delta v
	\end{bmatrix},
	\end{equation}
	where each entries of the Jacobian matrix are evaluated at critical points ($u_c$, $v_c$). The nature of the critical points are decided by the sign of the eigen values of the Jacobian matrix. The Jacobian matrix for the model is obtained as
	\begin{align} 
	\label{eqn:M2}
	\begin{bmatrix}
	3\nu-3(1-\nu-v) \hspace{0.1cm}& -3(1-u) \vspace{0.2cm}\\
	-3(v-\tilde{\xi_1}) & 3(\nu-u)
	\end{bmatrix}.
	\end{align}
	The eigenvalues of the matrix are obtained by diagonalizing the matrx. The critical points, eigenvalues and stability of the critical points are summariized in Table \ref{tab:table1}.
	\begin{table}
		\caption{\label{tab:table1}Critical points and eigenvalues of the phase space variables $u$ and $v$ and stability of the critical points.}
		\centering
		\begin{tabular}{|c|c|c|}
			\hline
			Critical points ($v_c,u_c$)   \hspace{0.15cm} & Eigen values  & Stability \\[2pt]
			\hline
			(0.9978, 0.0011)  \hspace{0.15cm} & (-2.9934, 2.9934)  & Saddle \\[2pt] 
			\hline
			(0.0011, 1)  \hspace{0.15cm} & (-2.9901, -2.9967 )  & Stable \\[2pt] 
			\hline
		\end{tabular}
	\end{table}
	The stability of the critical points is determined by the sign of the eigenvalues of the Jacobian matrix. The critical points are considered stable if all the eigenvalues of the Jacobian matrix are negative. Then, all the trajectories originating from the vicinity of the critical points converge to the critical points. The critical points are unstable if all the eigenvalues of the Jacobian matrix are positive. Then the trajectories at the vicinity of the critical points are all diverging irrespective of the initial conditions. If one of the eigenvalues of the Jacobian matrix is negative and the other is positive, then the critical points are said to be a saddle. The trajectories originating from the vicinity of the critical points may converge or diverge depending upon the initial conditions.
	
\begin{figure}[H]
		\centering
		\includegraphics[width=0.5\linewidth]{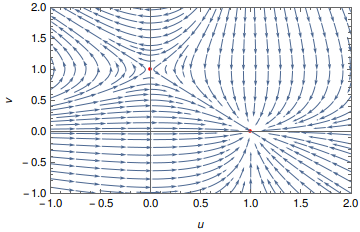}
		\caption{Phase space portrait of $u$ and $v$. Red dots are the critical points.}
		\label{fig:phase}
	\end{figure}

	The eigenvalues corresponding to the critical point $(0.9978, 0.0011)$ are negative and positive each, hence the point is a saddle one. Corresponding to this point, matter dominate over dark energy, which represent decelerated phase. Since the saddle is an unstable equilibrium point, the universe evolves further to attain a stable equilibrium state. On the other hand, the eigenvalues corresponding to the critical point $(0.0011, 1)$ are both negative; the critical point is a stable equilibrium point. Corresponding to this the dominant component is vacuum energy density, and the matter density is almost negligible, representing the stable final de Sitter phase. The phase space trajectory in the $u-v$ plane is shown in fig. (\ref{fig:phase}), from which, it is evident that the critical point $(0.0011,1)$ is a stable attractor because all the close trajectories converge to the point and $(0.9978, 0.0011)$ is an unstable saddle because some of the trajectories approach the critical point and others diverge.
	\section{GSL and Maximization of Entropy}
\begin{figure}[H]
		\centering
		\includegraphics[width=0.48\linewidth]{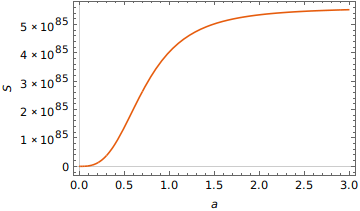}
		\caption{Evolution of total entropy with respect to redshift.}
		\label{fig:eT}
	\end{figure}

\begin{figure}[H]
		\centering
		\includegraphics[width=0.5\linewidth]{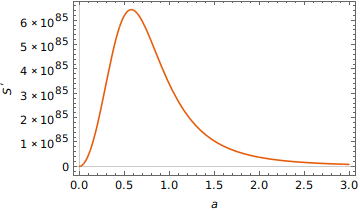}
		\caption{Rate of entropy change against scale factor and validity of GSL.}
		\label{fig:ed}
	\end{figure}

\begin{figure}[H]
		\centering
		\includegraphics[width=0.5\linewidth]{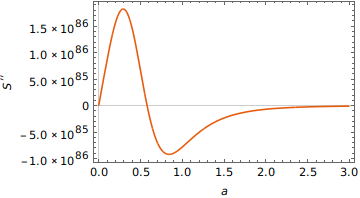}
		\caption{Evolution of $S''$ against scale factor for the best estimated model parameters.}
		\label{fig:edd}
	\end{figure}

	Here, we investigate the validity of the generalized second law (GSL) of thermodynamics. According to the GSL, the entropy of the cosmic components (matter, radiation and dark energy) inside the Hubble horizon plus the entropy associated with the Hubble horizon always increase with time,
	\begin{align}
	\label{eqn:e1}
	\frac{d}{dt}(S_H + S_m + S_{\Lambda})\geq 0,
	\end{align}
	where $S_H$ is the entropy associated with the Hubble Horizon or equivalently the apparent Horizon for a flat universe, $S_m$ is the entropy of the nonrelativistic matter and $S_{\Lambda}$ is the entropy contribution from the dark energy. Here, we are not considering the radiation entropy as it is not significant in the late phase. Gibb's equation connecting the total entropy with energy and pressure can be written as
	\begin{align}
	\label{eqn:e2}
	TdS = dE + pdV,
	\end{align}
	where $E=\rho V$ is the internal energy of the cosmic fluids having density $\rho$, $p$ is the pressure and V is the volume enclossed by the horizon. Here we assume that the temperature T associated with both matter and dark energy are equal because of their mutual interaction. We also assume that the system bounded by the Hubble horizon remains equilibrium so that the temperature distibution is uniform and its value equal to the temperature of the horizon\cite{karami2010generalized}. The Gibbons-Hawking temperature, $T=H/2\pi$ is a natural choice for the horizon temperature\cite{krishna2017holographic}. Using the integrability condition
	\begin{align}
	\label{eqn:e3}
	\frac{\partial^2S}{\partial T\partial V} = \frac{\partial^2S}{\partial V\partial T},
	\end{align}
	we obtain the relation connecting energy density and pressure as
	\begin{align}
	\label{eqn:e04}
	dp = \frac{\rho+p}{T}dT.
	\end{align}
	If we substitute (\ref{eqn:e04}) into (\ref{eqn:e2}), it follows that
	\begin{align}
	\label{eqn:e4}
	dS = d\left[\frac{(\rho + p)V}{T}+C\right],
	\end{align}
	where $C$ is the integration constant\cite{kolb2018early}. From eq. (\ref{eqn:e4}), it is evident that the vacuum energy density in the present model doesn't contribute to the entropy as $p_{\Lambda}=-\rho_{\Lambda}$. The non zero contribution to the entropy due to non relativistic matter up to an additive constant can be written as 
	\begin{align}
	\label{eqn:e5}
	S = \frac{(\rho_m + \text{p}_{\text{eff}})V}{T},
	\end{align}
	where $\rho_m$ is the matter density and $\text{p}_{\text{eff}}$ is the pressure due to bulk viscous matter. Substituting for $\rho_m$ given in eq. (\ref{eqn:rho3}), $\text{p}_{\text{eff}}$ given in eq. (\ref{eqn:p1}), $V=\frac{4\pi}{3H^3}$ and the Gibbon-Hawking temperature $T=\frac{H}{2\pi}$, we obtain the expression of matter entropy in a natural system of unit in  which $\hbar = c = k_B = 8\pi G =1$ as
	\begin{align}
	\label{eqn:e6}
	S_m = \frac{8\pi^2}{3H^4}(\rho_{m} + \text{p}_{\text{eff}}) ,
	\end{align}
	Substituting for $\rho_m$ and $\text{p}_{\text{eff}}$ from eq. (\ref{eqn:rho3}) and (\ref{eqn:p2}) respectevely, we obtain
	\begin{align}
	\label{eqn:e7}
	S_m = \frac{8\pi^2}{H^4}(\rho_{m_0} + \text{p}_{\text{eff}}^0) a^{-3(1-\nu-\tilde{\xi_1})}.
	\end{align}
	Acoording to Bekenstein, the horizon entropy is proportional to the area of the horizon. Here we consider Hubble horizon as the thermodynamic boundary. The area-entropy relation is
	\begin{align}
	\label{eqn:e8}
	S_H=\frac{k_B A}{4l_p^2},
	\end{align}
	where $k_B$ is the Boltzmann constant, $A=\frac{4\pi}{H^2}$ is the area of the Hubble horizon and $l_p=\sqrt{\frac{\hbar G}{c^3}}$ is the Planck length. Here we consider natural system of unit having $\hbar=c=k_B=8\pi G=1$, the eq. (\ref{eqn:e8}) reduces to
	\begin{align}
	\label{eqn:e9}
	S_H=\frac{8\pi^2}{H^2}.
	\end{align}
	The total entropy of the universe is
	\begin{align}
	\label{eqn:e10}
	S=S_H+S_m.
	\end{align}
	Sustituting the matter entropy from eq. (\ref{eqn:e7}) and horizon entropy from eq. (\ref{eqn:e9}), we obtain the expression for total enetropy as given in (\ref{eqn:e11}) and its evolution against the redshift is plotted in fig. \ref{fig:eT}
	\begin{align}
	\label{eqn:e11}
	S=\frac{8\pi^2}{H^2}\left(\frac{(\rho_{m_0} + \text{p}_{\text{eff}}^0) a^{-3(1-\nu-\tilde{\xi_1})}}{H^2}+1\right).
	\end{align}
	\noindent According to generalized conservation law (GSL), the total entropy of the universe never decreases, 
	\begin{align}
	\label{eqn:ed}
	\dot{S}=\dot{S}_H+\dot{S}_m\geq 0,
	\end{align}
	where the overdot represents the derivative with respect to cosmic time.
	To check the validity of the GSL in this model, it is convenient to change the variable from time to scale factor and evaluate the rate of change of entropy with respect to scale factor ($S'$), we obtain 
	\small
	\begin{equation}
	\label{eqn:ed1}
	S'=\frac{8\pi^2}{3H^4}\left[(\rho_{m_0}+\text{p}_{\text{eff}}^0)a^{-4+3(\nu+\tilde{\xi_1})}\right]\left[\frac{2(\rho_{m_0}+\text{p}_{\text{eff}}^0)a^{-3(1-\nu-\tilde{\xi_1})}}{H^2}+
	3(\nu+\tilde{\xi_1})\right].
	\end{equation}
	From eq. (\ref{eqn:ed1}), all the terms are positive so that $S'\geq 0$ and hence the GSL is satisfied. In the asymtotic limit $a\rightarrow\infty$, $S'\rightarrow 0$, it shows that the entropy is extremized and indicate that the end phase is an equilibrium state. The behaviour of $S'$ against scale factor is shown in fig. \ref{fig:ed}.
	\noindent The equilibrium state may be stable or unstable according to the sign of $S''$. The equilibrium state is stable if $S''< 0$ atleast in the far future in order to get the entropy bounded. The second derivative of entropy with respect to scale factor is obtained by diffrentiating the eq. (\ref{eqn:ed1})
	\begin{align}
	\label{eqn:edd}
	S''=\,& 8\pi^2\left[\frac{6(\rho_{m_0}+\text{p}_{\text{eff}}^0)^3a^{-10+9(\nu+\tilde{\xi_1})}}{H^8}\right.\nonumber\\&-\left.
	\frac{12(1-\nu-\tilde{\xi_1})(\rho_{m_0}+\text{p}_{\text{eff}}^0)^2a^{-7+6(\nu+\tilde{\xi_1})}}{H^6}\vphantom{\frac{1}{1}} \right]+
	\nonumber \\ &8\pi^2(\nu+\tilde{\xi_1})\left[\frac{2(\rho_{m_0}+\text{p}_{\text{eff}}^0)^2a^{-7+6(\nu+\tilde{\xi_1})}}{H^6}\right.\nonumber\\&-\left.\frac{3(\rho_{m_0}+\text{p}_{\text{eff}}^0)(1-\nu-\tilde{\xi_1})a^{-4+3(\nu+\tilde{\xi_1})}}{H^4}\right]
	\end{align}
	The evolution of $s''(a)$ for the estimated model parameters is shown in fig. \ref{fig:edd}. 
	\noindent From the figure, $S''>0$ in the early phase of evolution and made a transition to $S''<0$ in the recent past near the transition redshift, convexity condition is satisfied, and the entropy will be bounded. Therefore a thermodynamically stable equilibrium state will be achieved at the late stage. As $a\rightarrow\infty$ $S''\rightarrow 0$. Any system satisfying the extremization of entropy and convexity condition behaves like an ordinary macroscopic system. Therefore we can conclude that the evolution of the universe is like the evolution of an ordinary macroscopic system.
\section{Conclusions}
In this paper, we have studied the late accelerating expansion of the universe, considering the running vacuum energy and bulk viscous matter as the dominant components present in the universe. The vacuum energy evolves as the squares of the Hubble parameter, and it is motivated by the renormalization group equation of quantum field theory in curved spacetime geometry. The bulk viscosity coefficient of the matter is proportional to the expansion velocity so that the bulk viscous pressure also varies as the squares of the Hubble parameter. We have adopted Eckart's theory to incorporate bulk viscous pressure into the nonrelativistic matter. Even though the Eckart theory is noncausal and the equilibrium states in this theory are unstable, it is much easier than the causal Israel–Stewart theory, and many authors widely used it. The model successfully solves the cosmological constant problem and the coincidence problem because of the dynamical nature of vacuum energy density. We have used the chi-square minimization procedure to constrain the model parameters using the data set SN1a+CMB+BAO+OHD. The chi-square degrees of freedom obtained is close to one, stipulating a decent fit. The incorporation of bulk viscous matter with the running vacuum successfully constrains the bulk viscosity of the matter comparable to highly viscous materials observed on our planet. The estimated value of $\tilde{\xi_1}$ is $0.0011^{+0.0012}_{-0.0011}$, from which we can calculate the coefficient of bulk viscosity, $\xi = 1.316\times 10^5$ in SI unit, which falls in the predicted range $10^4 - 10^7$ for the matter-dominated bulk viscous models. The value is one to two orders of magnitude less than the value obtained using a class of bulk viscous matter-dominated models\cite{brevik2017viscous, mohan2017bulk, sasidharan2015bulk, sasidharan2018bayesian}. The parameter that is responsible for the vacuum dynamics is $\nu$, its estimated value is $0.0011^{+0.0020}_{-0.0011}$, which is almost similar to the value obtained by the running vacuum model\cite{tsiapi2019testing}. The non zero value of $\nu$ indicates the dynamical nature of vacuum energy density.

The Hubble parameter obtained is a decreasing function of the scale factor, and it attains a de Sitter evolution in the far future. We have estimated the age of the universe as 14 Gyr, which is slightly higher than the age-predicted by the $\Lambda$CDM model. However, it is a considerable improvement compared to matter-dominated bulk viscous models. The model fits the supernovae data and observational Hubble data for the best-estimated model parameters. The vacuum energy density and matter density shows distinct evolutionary behaviours. The nonrelativistic matter dominated the early universe so that the expansion of the universe was decelerating. The vacuum energy dominates matter in the near past, and hence the universe makes a transition to an accelerating expansion. The deceleration parameter plot clearly depicts the transition from matter-dominated decelerating phase to vacuum energy-dominated accelerating phase. The estimated transition redshift is 0.7306, which is close to the transition redshift obtained by the $\Lambda$CDM model.

We have performed the statefinder analysis to distinguish our model from the $\Lambda$CDM model. The evolution of the jerk parameter (r) and the snap parameter (s) in the r-s plane is confined to $r<1$ and $s>0$ until it reaches the $\Lambda$CDM model, shows the quintessence behaviour of our model. The present value of the r and s parameter is estimated as ($r_0$, $s_0$) = ($0.9972$, $0.0008$) while ($r$, $s$) = ($1$, $0$) is a fixed point for the $\Lambda$CDM model.  It shows a clear distinction between our model and the standard $\Lambda$CDM model.  

The dynamical system analysis is performed to understand the dynamical stability of the model. We have obtained two critical points; $(0.0011, 1)$ is a stable equilibrium point in the vacuum energy dominated phase, and $(0.9978, 0.0011)$ is a saddle corresponding to the unstable matter-dominated phase. The stable equilibrium state is achieved in the late de Sitter phase of the universe. 

The thermodynamical stability of the universe and the validity of the Generalized second law is verified by analyzing the progress of entropy in the expanding universe. The analysis shows that the decrease in matter entropy is balanced by the extensive increase in the horizon entropy so that the total entropy of the universe is ever increasing. Thus we can conclude that the universe satisfies the GSL throughout its evolution. The evolution of $S''$ is such that it asymptotically approaches zero from negative value, the convexity condition is satisfied; this indicates that the universe behaves like an ordinary macroscopic system.
	
\end{document}